\documentclass[amssymb, amsmath,cha]{nature}

\usepackage{graphicx}
\usepackage{dcolumn}
\usepackage{bm}
\usepackage{epstopdf}
\usepackage{color}
\usepackage{here}
\title{Optical Control of Young's Type Double-slit\\ Interferometer for Laser-induced Electron Emission\\ from a Nano-tip}

\author{Hirofumi Yanagisawa$^{1,2,3}$, Marcelo Ciappina$^4$, Christian Hafner$^5$, Johannes Sch\"{o}tz$^{2,3}$, J\"{u}rg Osterwalder$^6$, Matthias F. Kling$^{2,3}$}

\begin{document}

\maketitle

\begin{affiliations}
 \item Institute for Quantum Electronics, ETH Z\"{u}rich, CH-8093 Z\"{u}rich, Switzerland
 \item Max Planck Institute of Quantum Optics, D-85748 Garching, Germany
 \item Physics Department, Ludwig-Maximilians-Universit\"{a}t Munich, D-85748 Garching, Germany
 \item Institute of Physics of the ASCR, ELI-Beamlines, Na Slovance 2, 182 21 Prague, Czech Republic
 \item Laboratory for Electromagnetic Fields and Microwave Electronics, CH-8092 Z\"{u}rich, Switzerland
 \item Physik-Institut, Universit\"{a}t Z\"{u}rich, CH-8057 Z\"{u}rich, Switzerland
\end{affiliations}

\clearpage
\begin{abstract}
Interference experiments with electrons in a vacuum can illuminate both the quantum and the nanoscale nature of the underlying physics. An interference experiment requires two coherent waves, which can be generated by splitting a single coherent wave using a double slit. If the slit-edge separation is larger than the coherence width at the slit, no interference appears. Here we employed variations in surface barrier at the apex of a tungsten nano-tip as slits and achieved an optically controlled double slit, where the separation and opening-and-closing of the two slits can be controlled by respectively adjusting the intensity and polarization of ultrashort laser pulses. Using this technique, we have demonstrated interference between two electron waves emitted from the tip apex, where interference has never been observed prior to this technique because of the large slit-edge separation. Our findings pave the way towards simple time-resolved electron holography on e.g. molecular adsorbates employing just a nano-tip and a screen.
\end{abstract}

Electrons are particles, but they also have the characteristics of waves, which is the beauty and mystery of quantum mechanics \cite{mollenstedt54, jonsson61, merli76, oshima02, cho04,  bach13, ehberger15}. The wave nature of electrons is not only of fundamental interest for studying quantum phenomena but is also important in high-resolution electron microscopy \cite{gabor48, lin86}, scattering and imaging processes of high-resolution transmission microscopy \cite{rother09, coene92}, or electron holography that enables us to obtain vistas into the nanoscale world \cite{allard94, latychevskaia13} or even in the attosecond atomic realm \cite{huismans11}. Their wave nature can typically be observed by so-called Young's interference using a double slit \cite{jonsson61, oshima02, bach13}. In such an experiment, electrons pass through either side of a double slit and strike a detector that is some distance away from the slits. The intensity distribution at the detector will show an oscillating pattern that is not expected if the motion of an electron is described as that of a point travelling along a well-defined path \cite{feynmann65}. This phenomenon of interference can be understood by the wave nature: a single coherent wave is split into two coherent waves by the slits, and they interfere constructively or destructively depending on their relative phases, resulting in an oscillation in the signals. Because these two coherent waves must be created to observe the interference, the separation between the inner edges of the two slits must be shorter than the coherence width of the electrons at the slit. This condition typically requires elaborate mechanical designs with careful choices of materials for the electron optics and the double slit \cite{jonsson61, oshima02, bach13} (or, more generally, a beam splitter that includes a biprism \cite{mollenstedt54, merli76, cho04, ehberger15}). The coherence condition can typically be set up by steering an electron beam, controlling its magnifications or changing the mechanical configuration of the beam splitter \cite{cowley92}.

Here we have achieved optical control of the double-slit dimension using the simplest form of Young's interference, which is established via two electron beams from a nanometre-sized tip apex \cite{oshima02}. Applying high DC fields on the tip apex can drive electron tunnelling through the surface barrier, known as field emission \cite{gomer93}. The field emission current density depends exponentially on the integral across the surface potential-energy barrier of the quantity ${\rm [U-{E}_{n}]^{1/2}}$, where U is the electron potential energy, and ${\rm {E}_{n}}$ is its normal-energy level (energy level associated with motion normal to the emitter surface) \cite{murphy56, young59}. Therefore, a slight variation of the surface barrier dramatically changes the current. The surface potential energy is modulated by the local work function, which in turn varies with the crystallographic surface orientation along the curved tip surface. As a result, the emission sites become localized on the nanometre scale, and it is possible to establish two emission sites within the coherence width of the electrons inside the tip. Such a situation represents a Young's interference experiment where the coherent electron wave in the metal is split by a double slit upon field emission due to the modulated surface barriers. This kind of interference has never been observed other than at the tiny apex of a carbon nanotube (CNT) with a radius of 5\,nm \cite{oshima02}. Therefore, careful material and tip designs are necessary for observing the interference via this method, and controlling the double-slit properties is very difficult. An alternative candidate is believed to be a superconducting tip with macroscopically extended coherent electron waves \cite{oshima02}.

By using photo-assisted electron emission, and without changing the tip and materials, we could control the distance and opening-and-closing of the double slit, represented by the surface barriers, by tuning the parameters of 7\,fs laser pulses that induce the photoexcitation. We observed interference patterns from two electron beams from a comparatively large tungsten tip apex with a radius of curvature of approximately 100\,nm at room temperature. The underlying physics is derived by numerical modelling and shows that photo-excited electrons on the tip surface experience small slit distances, resulting in an interference that is visible between two adjacent laser-induced electron beams from the tungsten tip apex as schematically illustrated in Fig. 1(a) (see Method 1 for details of the experiments).

\begin{figure}[H]
\vspace{-30pt}
\begin{center}
\includegraphics[scale=0.15]{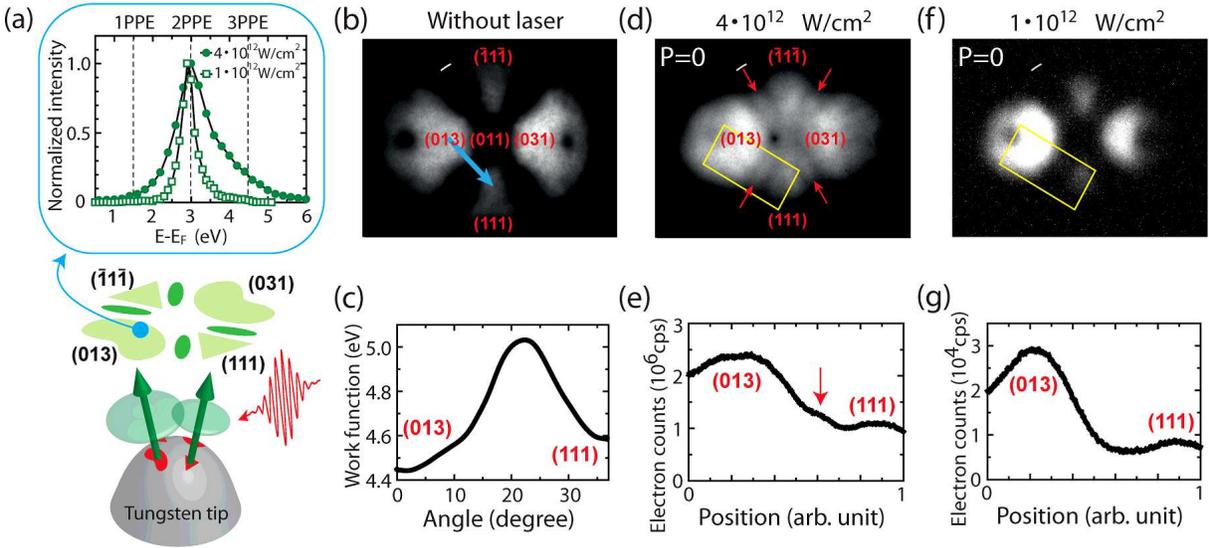}
\end{center}
\vspace{-20pt}
\caption{\label{fig:epsart}
{\small {\bf Conceptual diagram of the experimental setup and the electron emission patterns from the tungsten tip apex.} (a) Conceptual diagram of the experimental setup and observed interference. The inset shows energy spectra of laser-induced electron emission from the tungsten tip apex for two different laser intensities, $1 \cdot 10^{12}\,$W/cm$^{2}$ and $4 \cdot 10^{12}\,$W/cm$^{2}$. The electron emission from (310) type surface at the shadow side with respect to the laser propagation was measured. The spectra are normalized to their maximum values. (b) Field emission pattern without the laser. The extraction voltage between tip and counter electrode is 4600\,V (field estimated as around 5.5 V/nm). (c) The work function extracted from the electron signal variations along the blue line, indicated with an arrow in (b). The blue line represents the arc of the great circle of the tip hemisphere. (d) and (f) are laser-induced electron emission patterns taken at laser intensities of $4 \cdot 10^{12}\,$W/cm$^{2}$ and $1 \cdot 10^{12}\,$W/cm$^{2}$, respectively. Pink scale bars represent 10\,nm when the radius of curvature of the tip apex is 100\,nm, and they are arcs of a circle passing through centres of (001) type facets in the emission images on the detector plane. The extraction voltages were 1900\,V in both cases (field estimated as around 2.3 V/nm)}. The laser pulses propagate from right to left in the pictures. (e) and (g) intensity profiles of electron signals along the longer axis of the rectangles in (d) and (f); the electron signals are integrated over the shorter axis of the rectangles.\vspace{0pt}}
\end{figure}

\noindent {\bf Resutls and discussions}\\
\noindent {\bf Experimental}\\ Fig. 1(b) shows the typical field emission pattern from our tungsten tip apex oriented toward the [011] direction with four dominant emission sites: two from the (310) type surface facets and two from the (111) type. These emission sites are areas with lower surface barrier or work function. If electron emission originates purely from DC tunnelling, it is limited to only these four sites \cite{sato80}. In our laser-induced emission experiments, additional streaky patterns appear in the gaps between the (310) and (111) emission sites as shown in Fig. 1(d) and indicated with red arrows. The streaky pattern is quantitatively analyzed by integration over the shorter axis of the yellow rectangle in Fig. 1(d), yielding the curve in Fig. 1(e) with a small hump indicated by a red arrow. It should be noted that the width of each streaky pattern is approximately 10\,nm as indicated by a pink scale bar (representing 10\,nm).

Such streaky structures cannot be reproduced by simulating the laser-induced electron emission current with conventional Fowler-Nordheim (FN) theory \cite{murphy56, young59, yanagisawa09, yanagisawa10}. Previous studies showed that the emission mechanism in this regime is governed by a tunnelling emission from 2-photon photoexcitation (2PPE) or an emission over the top of the surface barrier from 3PPE \cite{yanagisawa11, yanagisawa12, yanagisawa16}, as also shown in the inset of Fig. 1(a). Note that the single-photon energy is approximate 1.5\,eV. Under these emission processes, the emission current depends on three factors: work function; local DC fields; and population of excited electrons on the tip apex \cite{yanagisawa11}. The last two factors are slowly-varying functions with respect to the positions on the tip apex for a smooth tip apex, which thereby cannot be expected to drive such streaky patterns. In contrast, the work function changes on a scale of nanometres on the tip apex. We extracted the work function map along the blue arrow in Fig. 1(b) (See the Method 2.1 for extracting the work function). However, it shows a simple monotonic increase towards the gap between the two emission sites as in Fig. 1(c), from which only a monotonic decrease of current can be expected towards the gap. As a matter of fact, the previous study based on FN theory has not shown the streaky emission patterns \cite{yanagisawa10}. Because FN theory does not take into account propagation of coherent electron waves in a vacuum, the streaky patterns are expected to be a result of interference of two electron beams in a vacuum.

As a clue for the underlying physics, the streaky patterns disappear when the laser intensity is reduced as shown in Figs. 1(f) and 1(g), which is consistent with some of our previous studies \cite{yanagisawa09, yanagisawa10}. Previous work showed the tunability of emission processes with laser intensity \cite{yanagisawa11}. When increasing the laser intensity, more electrons are emitted over the top of a barrier lowered by the Schottky effect via 3PPE processes as shown in the inset of Fig. 1(a) \cite{yanagisawa11}. Therefore, the appearance of the streaky feature would be associated with the advent of high energy electrons from 3PPE. In this work, our simulations based on the time-dependent Schr\"{o}dinger equation (TDSE) will confirm this hypothesis qualitatively.

\begin{figure}[t]
\begin{center}
\includegraphics[scale=0.13]{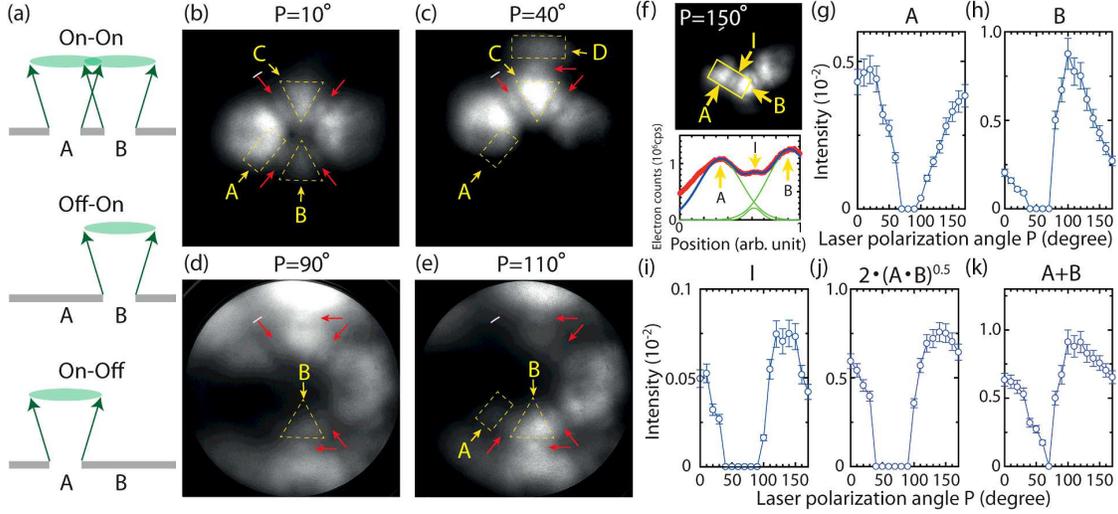}
\end{center}
\caption{\label{fig:epsart}
{\bf Polarization dependence of laser-induced electron emission patterns and their quantitative analysis.} (a) Conceptual diagram for interference when two emission sites, A, B are either on-on, off-on or on-off. (b)-(e) Laser-induced electron emission patterns for different laser polarization angles, P, which is defined by the angle between the tip axis and the polarization vector. Pink scale bars are the same as those used in Fig. 1. (f) Upper panel: laser-induced electron emission patterns at P=150$^{\circ}$. Lower panel: a line profile for the rectangular area using the same procedure as in Fig. 1(e) and 1(g). The three peaks in the line profile, assigned as A, B and I, are decomposed by Gaussian functions. (g)-(i) Plots of peak values of A, B and I as a function of polarization angle, P. (j) and (k) The same plots for $2 \cdot (A \cdot B)^{0.5}$ and A+B, respectively.\vspace{20pt}}
\end{figure}

How could we confirm the interference experimentally? As schematically sketched in Fig. 2(a), if one of the two electron beams is switched off, the interference patterns should disappear as is demonstrated using mechanical slits \cite{bach13}. We can realize such a situation using the site-selection technique found in our previous work where we can select the specific emission sites by changing the polarization angle of the laser \cite{yanagisawa09, yanagisawa10}. Some emission patterns for different polarization angles, P, are shown in Figs. 2(b) - 2(e). In all cases, the streaky patterns, indicated by red arrows, appear only when emission is present from two adjacent sites. For instance, looking at the sites A and B (each emission site is surrounded by dashed lines if it exists), the streaky patterns can be observed whenever A and B exist even if one of them is dim as seen in Figs. 2(b) and 2(e). However, the streaky pattern is not visible in case either A or B is absent as in Fig. 2(c) and 2(d). In another example, the interference pattern can be observed between C and D in Fig. 2(c), but it disappears when D is absent in Fig. 2(b). These observations are the same between any two adjacent sites for our conditions of detection efficiency and screen resolution.

Further quantitative analysis indicates the strong signature of the interference. As shown in Fig. 2(f), we evaluated the signal profile for the rectangle area as we did in Fig. 1(c) and 1(d). The obtained profile was decomposed using Gaussian functions for A, B and I sites defined in Fig. 2(f), and the peak values of the Gaussians were divided by the total count rate of the rectangle area. The errors of peak values arise due to the uncertainty of the peak position of the Gaussians, which are estimated to be 10 \% at most. In addition, when the signal level is too low and it is hard to assign Gaussian functions, we set the peak values to zero. Then they were plotted as a function of laser polarization angle for A, B and I in Fig. 2(g) - 2(i), respectively.These three sets clearly show that I becomes 0 when either A or B is zero. We also have inspected the quantity $2 \cdot (A \cdot B)^{0.5}$ as in Fig. 2(j) because the intensity of interference should follow the product term of the amplitude of two wave functions \cite{feynmann65}. Indeed, the quantity varies similarly with that of I as in Fig. 2(i). In contrast, the quantity, A+B behaves rather differently from I. Hence, these data sets strongly indicate that the streaky pattern is not a phenomenon driven by a simple sum of the two emissions but an interference phenomenon. We have checked other site combinations, which confirmed the observations and conclusion. 

\begin{figure}[p]
\begin{center}
\vspace{-20pt}
\includegraphics[scale=0.2]{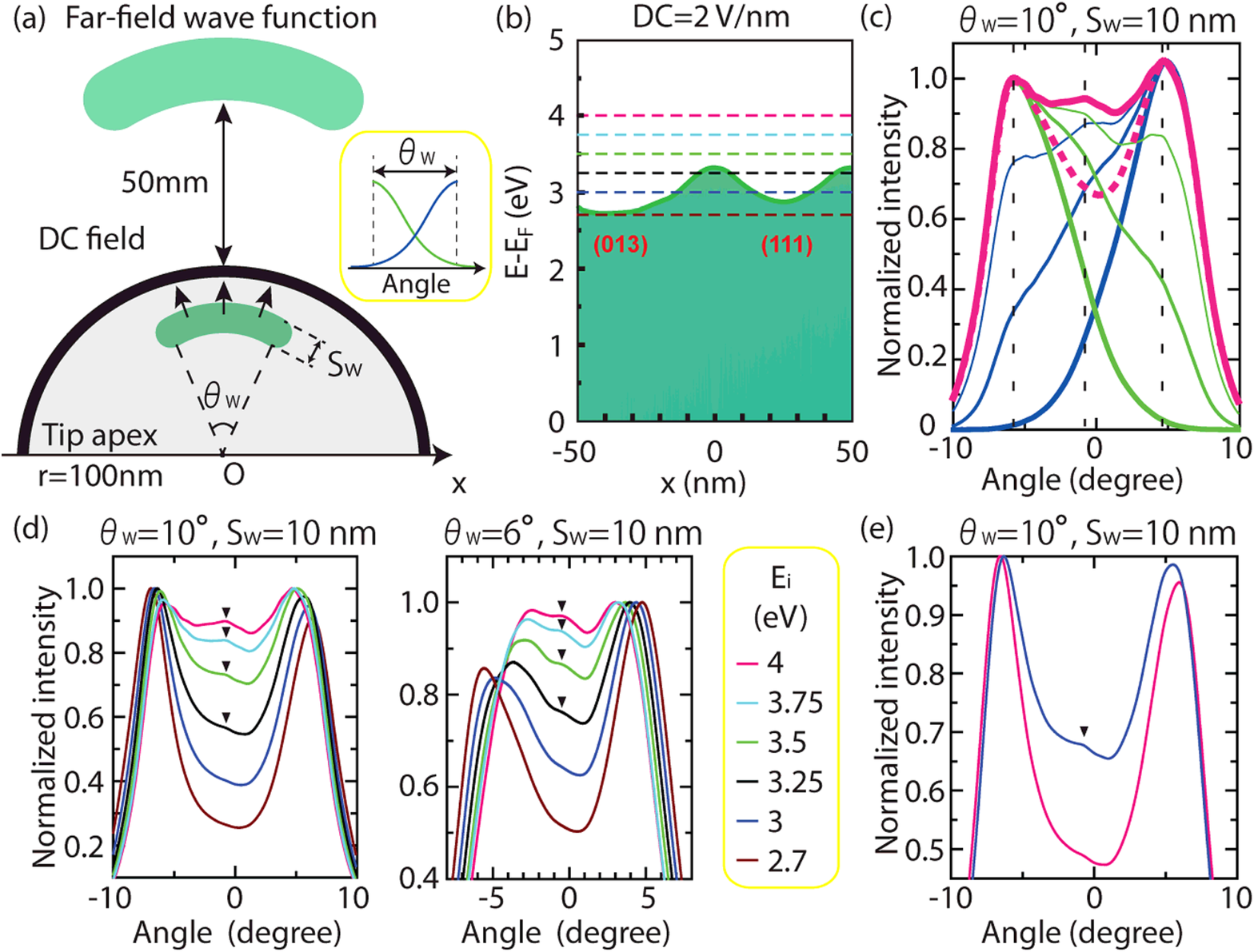}
\end{center}
\vspace{-20pt}
\caption{\label{fig:epsart}
{\bf Far-field electron intensity distributions simulated by TDSE.} (a) Concept of our simulation model. The inset shows half Gaussian functions to attenuate the initial wave functions for simulations in (c). See text for details. (b) Electron potential landscape under the DC field of 2V/nm, where the energy is measured with respect to the Fermi level, $E_{F}$. (c) Far-field intensity distributions as a function of angle with respect to the center axis (thick solid line). The initial radial energy $E_{i}$ is 4eV defined with respect to the Fermi energy. Other curves are the results of the attenuated initial wave functions. See the text for details. (d) Far-field intensity distributions for various initial energies, which are indicated in the inset and also (b) by dashed lines. The same color codes were used in Fig. 2(b) and 2(d) to indicate initial energies. (e) Energy-integrated far-field intensity distributions reconstructed by simulated intensity distributions with weight extracted from energy spectra. The blue line is for laser intensity of $4 \cdot 10^{12}\,$W/cm$^{2}$ and the pink line for $1 \cdot 10^{12}$\,W/cm$^{2}$
\vspace{20pt}}
\end{figure}

\clearpage
\noindent {\bf Simulations}\\ Our experimental observation can indeed be qualitatively reproduced by simulations with simple assumptions. In the simulation, we have generated a coherent spherical wave packet inside the tip apex and let it propagate under DC fields up to 50\,mm away from the apex, where the detector is placed, by solving a two-dimensional TDSE \cite{endoh93}; the situation is schematically drawn in Fig. 3(a) (see Methods 2.2 - 2.4 for more details on the simulations). We have thereby obtained far-field wave functions and from them the electron density distributions. The dimensions of the initial wave packet are defined by two parameters, cone angle $\theta_{w}$ and width $S_{w}$. The image potential was employed to calculate the surface barrier landscape, the heights of which are determined by the work function landscape obtained from Fig. 1(b). Fig. 3(b) shows the surface potential energy barrier in our simulations for DC fields.

We integrated the intensity of the simulated far-field wave function over the radial direction and plotted them as a function of angle with respect to the center axis with the coordinate system placed in the center of the tip apex. The thereby simulated far-field intensity distributions successfully reproduces the interference peak as shown by the thick pink line in Fig. 3(c). There are three peaks as their positions indicated by thin vertical dashed lines. The peaks at both sides are from the two emission sites and their interference yielding an extra peak in the center, which disappears with switching off either of the emission sites as we observed experimentally. To switch off either of the sites, we multiplied the initial wave packet amplitude with the right (left) half of a Gaussian function centered at the leftmost (rightmost) end of the region defined by $\theta_{w}$ as shown by green (blue) lines in the inset of Fig. 3(a), and then we computed the far-field intensity distribution for different half-width-at-half-maximum (HWHM) of the Gaussian function; the results are shown by green (blue) lines in Fig. 3(c). The thicker the lines the narrower the Gaussian functions (HWHM are 24, 12 or 6$^{\circ}$ from the thinnest line). The results show that the two peaks in the attenuated region disappear concomitantly, which is the same behavior we observed in Fig. 2. The narrowest Gaussian distributions give only one emission site on either side; the sum of these spectra shows no interference peak at the center as indicated by a thick pink dashed line.

Further simulations revealed the energy dependence of the interference peak for six different initial energies as indicated by dashed lines in Fig. 3(b). The resulting far-field intensity distributions for $\theta_{w} = 10^{\circ}$ and $S_{w} = 10$\,nm are shown in Fig. 3(d). Clearly, the interference peak evolves with increasing initial energies. At $E_{i}$ = 2.7\,eV, the interference peak is barely seen. But at $E_{i}$=3.25\,eV which is around the top of the potential energy barrier, the interference peak appears. The energy dependence can also be seen for $\theta_{w}=6^{\circ}$ as shown in the next panel. Under this condition, the transverse extent of the coherent wave packet is around 10\,nm at the tip surface, which is almost the same as the coherence width of the electron in tungsten at room temperature \cite{cho04, ehberger15}, implying that interference may be observed for our conditions. Note that the coherence width of the excited electrons has been shown to be similar to that of ground states \cite{ehberger15}. For even shorter transverse extent of around 7\,nm, the interference peak disappears (See Fig. S1 in Supplementary Information). Additionally, the far-field intensity distributions stay similar for temporally broadened pulses (See Fig. S2 in Supplementary Information), which implies that the interference is not peculiar to the ultrashort pulse. Finally, the obtained energy dependent far-field distributions are integrated with weights extracted from the measured energy spectra taken by the laser with intensities of $1 \cdot 10^{12}$\,W/cm$^{2}$ and $4 \cdot 10^{12}$\,W/cm$^{2}$ (See the Methods 2.5 for details) as indicated by pink and blue lines in Fig. 3(e), respectively. The growth of the interference peak for higher intensities is clearly visible.

\begin{figure}[p]
\begin{center}
\vspace{-25pt}
\includegraphics[scale=0.13]{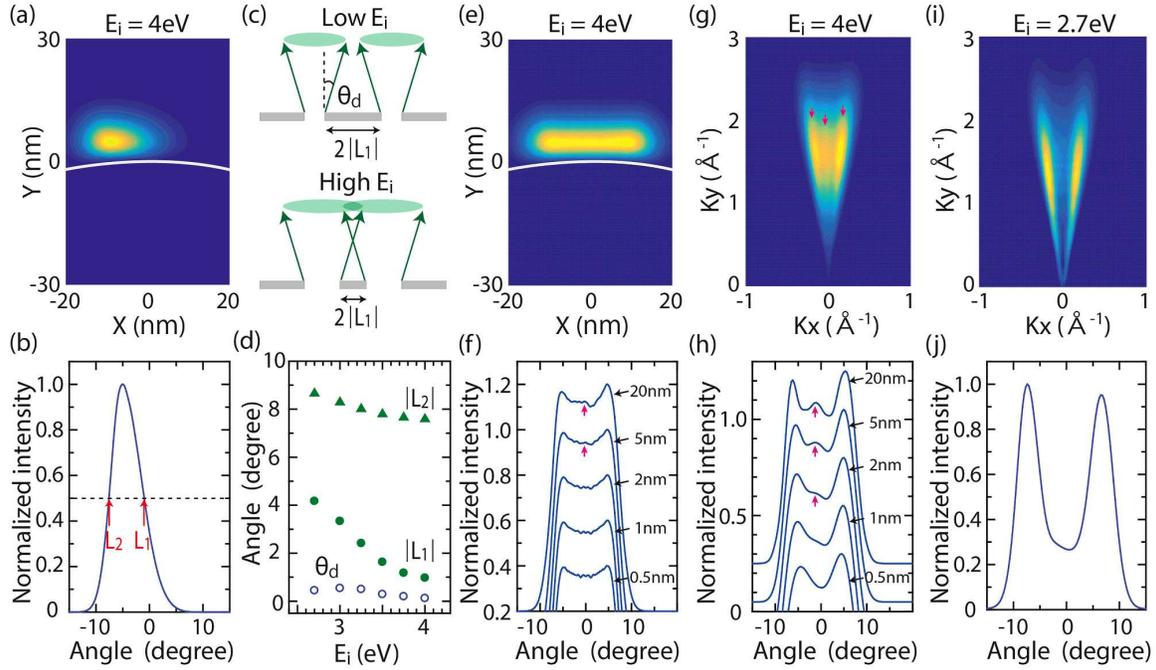}
\vspace{-25pt}
\end{center}
\caption{\label{fig:epsart}
{\small {\bf Near-field electron intensity and momentum distributions simulated by TDSE and analysis of electron beam parameters.} (a) and (e) Near-field electron intensity distribution when the maximum amplitude of the wave function along the center axis reaches a point 5\,nm away from the tip apex. (b) and (f) show their intensity profiles. The profile is made by integrating intensities along the radial direction and plotting them as a function of angle with respect to the center axis. $L_{1}$ and $L_{2}$ are angles at half maximum. (c) Conceptual diagram of changes of effective slit distance 2$\left| L_{1} \right|$ depending on the initial electron energy. Beam divergence $\theta_{d}$ is defined by the difference in angles between $L_{1}$ and the corresponding angle in the far-field intensity profile. (d) Variation of $\left| L_{1} \right|$, $\left| L_{2} \right|$ and $\theta_{d}$ as a function of initial energy $E_{i}$. (g) and (i) Momentum distributions of wave functions at 5\,nm away from the tip surface for initial energies of 4\,eV and 2.7\,eV, respectively. (h) and (j) are the intensity profiles of (g) and (i) along the polar direction. (f) and (h) also show profiles of electron intensity and momentum distributions, respectively, when the wave function reaches 20\,nm, 2\,nm, 1\,nm, and 0.5\,nm from the surface.
}}
\end{figure}

We would now like to gain an intuitive idea on which factor drives the interference in the context of Young's interference experiments. Unlike the original Young's double slit, the double slits in this method do not completely block electrons impinging on the surface barrier between two slits. This is because electrons can either leak through the barrier via tunnelling or be emitted over the barrier via photoemission, as discussed in the inset of Fig. 1(a). Therefore, the present implementation does not directly represent Young's double slits in this sense, but a single electron wave experiencing lateral modulation in its amplitude due to reflection at the surface barrier, resulting in two apparent beams. As already shown in Fig. 3(c), the existence of two beams is necessary for the advent of the interference. Hence, we regard this system as analogous to a Young's double-slit interferometer. In order to evaluate the dimensions of the double slits, the intensity profile was inspected when an electron wave is emitted from only one of the emission sites, which is the same condition as for the thickest green line in Fig. 3(c). First, we calculated the near-field wave function at 5\,nm distance away from the tip apex, and then computed radially integrated intensity profiles along the polar direction as in Figs. 4(a) and 4(b), respectively. By evaluating the positions at the half maximum of the profile, $L_{1}$ and $L_{2}$ as in Fig. 4(b), we deduced how the effective slit size changed with increasing initial energy. Especially, because $L_{1}$ indicates the right side edge of the emitted wave functions in near field, $L_{1}$ roughly tells us the effective distance between the two slits as indicated in Fig. 4(c). Second, we have investigated the beam divergence, $\theta_{d}$, by comparing the intensity profiles between near- and far- field wave functions as also indicated in Fig. 4(c). All these values are plotted as a function of the initial electron energy in Fig. 4(d). The results show that $L_{2}$ and the divergence change slightly but $L_{1}$ becomes significantly smaller at higher energies. Hence, we concluded that the change of the effective slit distance drives the interference for higher initial energies as schematically shown in Fig. 4(c). This tendency can be intuitively understood by Fig. 3(b), where the barrier width that excited electrons feel becomes narrower with increasing electron energy. Note that the top of the potential energy barrier is situated around 3.25eV as shown in Fig. 3(b). Classically, any electron with normal energy above the barrier height can be emitted over the barrier. Quantum mechanically, however, such electrons will still be scattered at the barrier and transmission rate is not unity upon emission just above the top of the barrier \cite{murphy56, cutler64}. This can be seen from the fact that $L_{1}$ changes its value even above 3.25\,eV. In addition, as seen in the near-field wave function at 5\,nm distance away from the tip apex (Fig. 4(e)) and radially integrated intensity profiles (Fig. 4(f)), two beams are generated for the initial energy of 4\,eV with conditions equivalent to those for the thick pink line in Fig. 3(c). Hence the surface potential energy barrier works as a beam splitter at least up to 4\,eV.

Finally, we would like to point out that we are working under emission conditions of more than 1 electron per pulse, which implies that space charge effects may need to be considered. In fact, we observed slight broadening of energy spectra due to space charge effect around $4 \cdot 10^{12}$\,W/cm$^{2}$ in our previous work \cite{yanagisawa16}. However, we consider that the interference phenomenon will not be significantly affected by space charge effects because the interference established in the close vicinity of the tip surface according to our simulations, as shown in Fig. 4(g). The figure shows the momentum distribution of the wave packet at 5\,nm away from the tip surface for the initial energy of 4\,eV with conditions equivalent to those for the thick pink line in Fig. 3(c). The data displays a center component due to the interference, which is also clearly seen in the intensity profile in Fig. 4(h) where the radially integrated intensity profiles of momentum distributions are plotted along the polar direction. In contrast, the center component disappears for the initial energy of 2.7\,eV as evident in Figs. 4(i) and 4(j). The data are consistent with the energy dependence of the far-field distribution in Fig. 3(d). The simulations further revealed that the interference peak evolves as the wave packet moves away from the surface on the scale of a couple of nanometers, as shown in Fig. 4(h). The interference peak appears around 2\,nm away from the surface. This is consistent with claims in previous work \cite{oshima02}. Note that the interference peaks can also be seen in the near-field wave functions in Fig. 4(f). The space charge effects become significant by accumulating Coulomb forces from the other electrons while the electrons are propagating in the vacuum, especially in the first 10-100$\mu m$ from the surface \cite{leuenberger11}, and they should be weak in the close vicinity of the surface due to Coulomb forces of opposite sign from the image charges of the emitted electrons. Hence, the observed interference should not to be affected by space charge effects.

\noindent {\bf Conclusion}\\
In this study, we have observed interference patterns between two electron beams induced by intense laser pulses from a tungsten tip apex; such interference is typically not visible due to the large separation of emission sites. TDSE simulations indicate that the observed interference is conceptually the same as Young's interference. We have found that photo-excited electrons effectively reduce the slit-edge separation and thus are responsible for the interference. Our findings provide a new degree of freedom to control the appearance of interference that can potentially be applied to other metal nanostructures. The tungsten tip used in this experiment is a suitable substrate for depositing molecules \cite{melmed58}. By depositing molecules on the tip's apex, we should be able to perform electron holography without constructing complicated electron optics and mechanical double slits as previously demonstrated using CNTs \cite{martin07}. This holography can survive space charge effects at high intensities because the interference occurs in the vicinity of the tip apex. Moreover, radiation damage to molecules due to energy deposition should be reduced because incident electron energy at the molecules is supposed to be very low, around 0.5 eV \cite{egerton04}. (On the downside, the spatial resolution would be in a range of a few nanometres because of the long wavelength.) More importantly, the demonstrated polarization dependence should realize two coherent electron waves from two consecutive emission sites with a relative time delay using two laser beams, which would enable us to experimentally analyze the temporal structure of electron wave-packets scattered at adsorbate molecules by investigating the interference pattern over changes to their delay time. This method would achieve time-resolved electron holography, possibly with attosecond time resolution because the interference should be sensitive to the relative phase between reference and scattered wave-packets.

\newpage
\noindent {\bf Methods}\\
\noindent {\bf 1. Experimental.}\\
A tungsten tip is mounted inside an ultra-high vacuum chamber ($9 \cdot 10^{-11}$\,mbar). Laser pulses are generated by an oscillator (centre wavelength: 830\,nm; repetition rate: 80\,MHz; pulse duration: 7\,fs) and introduced into the vacuum chamber. Using spectral phase interferometry for direct electric-field reconstruction (SPIDER) outside of the vacuum chamber, we confirmed that the pulse width could reach 7\,fs. A parabolic mirror in the chamber focuses the laser to approximately 3.5\,$\mu$m diameter ($e^{-2}$ radius) onto the tip apex. The tip was mounted on a 5-axis stage controlling three Cartesian coordinate positions $x, y, z$, as well as a tilt angle $\vartheta$ and an azimuthal angle $\varphi$ around the tip axis. Linearly polarized laser light was used, and the polarization vector adapted with a $\lambda/2$ plate. To observe the electron emission patterns from the tip, a two-dimensional detector was used (OCI-LEED detector). To measure the energy spectra of emitted electron, a hemisphereical analyzer was used (VG: CLAM2). The emission site to be measured was selected using a pinhole plate; details are described elsewhere \cite{yanagisawa11, yanagisawa12, yanagisawa16}. Cleanliness of the tip apex can be assessed from the emission patterns. If the surface is clean, we can observe the emission pattern shown in Fig. 1(b). A clean tip surface was prepared by flash-annealing the tip. Since the tip apex can be quickly contaminated even under ultra-high vacuum conditions, all the measurements were done within 15 min after sample heating.

\noindent {\bf 2. Simulation of far-field interference patterns.}\\
\noindent {\bf 2.1 Extracting the work functions.}\\
The work functions $\Phi$ along the blue arrow in Fig. 1(b) were extracted using the same procedure as in our previous work \cite{yanagisawa09, yanagisawa10}. Because the field emission current density can be described by $\Phi$ and DC fields $F$ in FN theory, $\Phi$ can be extracted from the experimentally obtained emission current if $F$ is given. A relative $F$ distribution on the tip apex was generated by OpenMaXwell \cite{openmax}. The absolute values were adjusted multiplying the $F$ distribution with a constant factor, and the resulting $\Phi$ map was compared to known values for several surface facets of tungsten. We have obtained 4.55\,eV, 4.45\,eV and 5.25\,eV for the (001), (310) and (011) surfaces, respectively. The resulting maximum DC field was 5.5V/nm. These values are in fair agreement with previous data \cite{michaelson77, mendenhall34}. Minor uncertainties in the work functions and DC fields do not affect the main conclusions in this article.

\noindent {\bf 2.2 Wave packet propagation by TDSE.}\\
We have simulated the temporal evolution of the created electron wave packets inside the tip apex by solving the TDSE in a two-dimensional system. The tip apex is assumed as hemispherical, and the radius of curvature is 100\,nm as shown in Fig. 3(a). The TDSE is solved using the pseudo-spectral method \cite{devries94}; our code successfully reproduced previous results in Ref. [\cite{endoh93}]. The time step is $2^{-18}$\,s. The propagation step size along the $x$ and $y$ axis is 0.98\,\AA \hspace{1pt} and 0.34\,\AA, respectively, where the $x$-axis is the horizontal axis and the $y$-axis the longitudinal axis in Fig. 3(a). It is difficult to compute the propagation of the electron wave up to 50\,mm away from the tip apex, where our detector is located, by just solving the TDSE. Therefore, after the electron kinetic energies reach 100\,eV, we used a propagator under constant DC fields. For the case of solving the TDSE under constant DC fields, an analytical formula can be obtained, and one can simulate wave functions after any time interval \cite{lukes69}. This allowed us to obtain the far-field wave functions. The details of the potential energy landscape are described below.

\noindent {\bf 2.3 Potential energy landscape.}\\
The potential energy inside the tip is set to be constant, assuming a free electron model. The surface barrier is modeled by an image potential. The height of the surface barrier with respect to the potential energy inside the tip is the Fermi energy (9.2\,eV \cite{islam09}) plus the work function. The work functions in Fig. 1(c) were used in the simulation. Here, the vacuum level distant from the emitter was taken as equal to ${\rm E_{F}+\phi_{min}}$, where ${\rm \phi_{min}}$ is the lowest value of local work-function used in the simulations. We assume that the differences in the work functions are converging to zero by following $1/d$ after a threshold value 1\,nm from the tip apex where $d$ is the distance from the surface. The threshold value of 1\,nm was used because previous work shows the maximum of the work function around 1\,nm from the surface \cite{fall99}. The choice of the threshold value, however, does not affect the main outcome; we have tested this up to a threshold of 50\,nm. We also applied a DC voltage between the tip apex and the detector. The DC field distributions were assumed spherically symmetric. Along the radial direction, we used the same DC potential distribution determined for the tip apex in the previous work \cite{yanagisawa16}. Because of computation limitations in TDSE described above, the DC field is set to be constant after the electron energy reaches 100\,eV, which is approximately 100\,nm away from the tip in our simulations. The value of a constant DC field is determined such that the final kinetic energy becomes approximately 2500\,eV.

\noindent {\bf 2.4 Initial wave function.}\\
As shown in Fig. 3(a), we have created a spherical wave packet inside the tip apex. The center of the spherical wave coincides with the center of the hemisphere. The amplitude of the spherical wave is constant along the polar direction within the cone angle $\theta_{w}$. Because the abrupt truncation of the wave function will cause energy broadening in their energy distribution, the wave packets over $\theta_{w}$ are truncated by following a Gaussian distribution; their half width at half maximum is 3.5\,nm. Along the radial direction, a Gaussian distribution is also applied, and $S_{w}$ in Fig. 3(a) corresponds to the full width at half maximum. The wavenumber of the spherical wave is determined by the initial electron energy.

\noindent {\bf 2.5 Reconstruction of energy-integrated far-field intensity distribution from the measured energy spectra.}\\
Reconstruction of energy-integrated far-field intensity distributions from the measured energy spectra in Fig. 3(e) was done in the following steps. Since the energy spectra in the inset in Fig. 1(a) is measured for (310) type facet, the energy spectra cannot be directly used as weight. Hence, first we have calculated electron distribution functions of the excited electrons by dividing the spectrum by the expected transmission probability assuming that the work function is 4.45\,eV and the DC field is 3\,V/nm, following previous work to extract the values \cite{yanagisawa11}. The thus obtained electron distribution functions were used as weights assuming that the distribution functions are homogeneous between (310) and (111) type facets. Energy-integrated far-field intensity distributions for a laser intensity of $4 \cdot 10^{12}$\,W/cm$^{2}$ ($1 \cdot 10^{12}$\,W/cm$^{2}$) are reconstructed by integrating the far-field intensity distributions with the initial energies from 1.6\,eV to 6\,eV (5\,eV) in 0.2\,eV (0.1\,eV) steps, multiplying with the obtained weights. Here we should emphasize a discrepancy in our theory. In our TDSE simulations, we considered only electrons striking the surface with normal momentum. As described in the previous work \cite{young59}, however, the calculation of emission current density needs to consider all the electrons impinging on the surface from all directions. This discrepancy in our theory should not affect the conclusion of the present work, but more sophisticated treatment is required for the quantitative analysis of the interference peak.

\clearpage

\begin{addendum}
\item This work was supported by the Swiss National Science Foundation through the {\it Ambizione} (Grant No PZ00P2\_131701) and the {\it NCCR MUST}, Kazato Research Foundation, the National Research Foundation of Korea through Global Research Laboratory Program [Grant No 2009-00439] and Max Planck POSTECH/KOREA Research Initiative Program [Grant No 2016K1A4A4A01922028], the DFG via SPP1840, the EU via the grant ATTOCO (Grant No 307203) and the project ELI–Extreme Light Infrastructure–phase 2 (Project No. CZ.02.1.01/0.0/0.0/15\_008/0000162) from European Regional Development Fund. We thank Prof. K. Watanabe and Dr. Vladislav S. Yakovlev for fruitful discussions.

\item[Author contributions] H.Y designed the study. H.Y constructed the experimental setup and carried out the experiments. H.Y performed the simulations with the support of M. C. and C. H. The manuscript was written by H. Y with the support of M. F. K. All the authors discussed the results and commented on the manuscript.

\item[Correspondence] Correspondence and requests for materials should be addressed to \\H.Y. (hirofumi.yanagisawa@mpq.mpg.de).
\end{addendum}

\begin{thebibliography}{99}


\bibitem{mollenstedt54} M\"{o}llenstedt, G. \& D\"{u}ker, H. Fresnelscher interferenzversuch mit einem biprisma f\"{u}r elektronenwellen. Naturwissenschaften, {\bf 42}, 41 (1954).

\bibitem{jonsson61} J\"{o}nsson, C. Elektroneninterferenzen an mehreren künstlich hergestellten Feinspalten. Zeitschrift für Physik, {\bf 161}, 454–474 (1961).

\bibitem{merli76} Merli, P. G., Missiroli, G. F. \& Pozzi, G. On the statistical aspect of electron interference phenomena. Am. J. Phys. {\bf 44}, 306 (1976).

\bibitem{oshima02}Oshima, C. {\it et al.} Young's Interference of Electrons in Field Emission Patterns. Phys. Rev. Lett. {\bf 88}, 038301 (2002).

\bibitem{cho04}Cho, B. {\it et al.} Quantitative Evaluation of Spatial Coherence of the Electron Beam from Low Temperature Field Emitters. Phys. Rev. Lett. {\bf 92}, 246103 (2004).

\bibitem{bach13}Bach, R. {\it et al.} Controlled double-slit electron diffraction. New J. Phys. {\bf 15}, 033018 (2013).

\bibitem{ehberger15}Ehberger, D. {\it et al.} Highly Coherent Electron Beam from a Laser-Triggered Tungsten Needle Tip. Phys. Rev. Lett. {\bf 114}, 227601 (2015)

\bibitem{gabor48}Gabor, D. A new microscopic principle, Nature {\bf 4098}, 777 (1948).

\bibitem{lin86}Lin, J. A. \& Cowley, J. M. Reconstruction from in-line electron holograms by digital processing. Ultramicroscopy {\bf 19}, 179-190 (1986).

\bibitem{rother09} Rother, A. \& Scheerschmidt, K. Relativistic effects in elastic scattering of electrons in TEM. Ultramicroscopy {\bf 109}, 154-160 (2009).

\bibitem{coene92} Coene, W. {\it et. al.} Retrieval through Focus Variation for Ultra-Resolution in Field-Emission Transmission Electron Microscopy. Phys. Rev. Lett. {\bf 69}, 3743-3746 (1992).

\bibitem{allard94}Allard, L. F. {\it et al.} Electron holography reveals the internal structure of palladium nano-particles. J. Mat. Sci. {\bf 29}, 5612-5614 (1994).

\bibitem{latychevskaia13}Latychevskaia, T., Longchamp, J.-N., Escher, C. \& Fink, H.-W. Coherent diffraction and holographic imaging of individual biomolecules using low-energy electrons. Present and Future Methods for Biomolecular Crystallography, Springer (2013).

\bibitem{huismans11} Huismans, Y. {\it et al.} Time-resolved holography with photoelectrons. Science {\bf 331}, 61-64 (2011). 

\bibitem{feynmann65} Feynman, R. P. \& Leighton, R. B., Sands, M. The Feynman Lectures on Physics, Vol. 3. (Addison-Wesley, 1965).

\bibitem{cowley92} Cowley, J. M. Twenty forms of electron holography. Ultramicroscopy {\bf 41}, 335-348 (1992).

\bibitem{haine52} Haine, M. E. \& Mulvey, T. The formation of the diffraction image with electrons in the Gabor diffraction microscope. J. Opt. Soc. Am. {\bf 42}, 763 (1952).


\bibitem{gomer93}Gomer, R. Field Emission and Field Ionization (American Institute of Physics, New York, 1993).

\bibitem{murphy56}Murphy, E. L. \& Good, R. H. Jr. Thermionic Emission, Field Emission, and the Transition Region. Phys. Rev. {\bf 102}, 1464 (1956).

\bibitem{young59}Young, R. D. Theoretical Total-Energy Distribution of Field-Emitted Electrons. Phys. Rev. {\bf 113}, 110 (1959).

\bibitem{sato80}Sato, M. Gas Adsorption on Tungsten Exposed to a Mixture of Nitrogen and Oxygen. Phys. Rev. Lett. {\bf 45}, 1856 (1980).

\bibitem{yanagisawa09}Yanagisawa, H. {\it et al.} Optical Control of Field-Emission Sites by Femtosecond Laser Pulses. Phys. Rev. Lett. {\bf 103}, 257603 (2009).

\bibitem{yanagisawa10}Yanagisawa, H. {\it et al.} Laser-induced field emission from a tungsten tip: Optical control of emission sites and the emission process. Phys. Rev. B {\bf 81}, 115429 (2010).

\bibitem{yanagisawa11}Yanagisawa, H. {\it et al.} Energy Distribution Curves of Ultrafast Laser-Induced Field Emission and Their Implications for Electron Dynamics. Phys. Rev. Lett. {\bf 107}, 087601 (2011).

\bibitem{yanagisawa12}Yanagisawa, H. Site-selective field emission source by femtosecond laser pulses and its emission mechanism. Ann. der Phys. {\bf 525}, 126-134 (2012).

\bibitem{yanagisawa16}Yanagisawa, H. {\it et al.} Delayed electron emission in strong-field driven tunnelling from a metallic nanotip in the multi-electron regime. Sci. Rep. {\bf 6}, 35877 (2016).

\bibitem{leuenberger11}Leuenberger, D. {\it et al.} Disentanglement of Electron Dynamics and Space-charge Effects in Time-resolved Photoemission from h-BN/Ni(111). Physical Review B {\bf 84}, 125107 (2011).

\bibitem{melmed58} Melmed, A. J. \& M\"{u}ller, E. W. Study of Molecular Patterns in the Field Emission Microscope. J. Chem. Phys. {\bf 29}, 1037 (1958).

\bibitem{martin07}Martin, G. L. \& Schwoebel, P. R. Field electron emission images of multi-walled carbon nanotubes. Surface Science {\bf 601}, 1521 (2007). 

\bibitem{egerton04}Egerton, R. F., Li, P. \& Malac, M. Radiation damage in the TEM and SEM. Micron {\bf 35} 399-409 (2004).

\bibitem{openmax}Hafner, C., OpenMaXwell. http://openmax.ethz.ch/ (2014).

\bibitem{michaelson77}Michaelson, H. B. The work function of the elements and its periodicity. J. Appl. Phys. {\bf 48}, 4729 (1977).

\bibitem{mendenhall34}Mendenhall, C. E. \& DeVoe, C. F. The Photoelectric Work Functions of the 211 and 310 Planes of Tungsten. Phys. Rev. {\bf 51}, 346 (1937).

\bibitem{devries94} DeVries, P. L. A First Course in Computational Physics. (John Wiley \& Sons, Inc., 1994).



\bibitem{endoh93} Endoh, A. {\it et al.} Time-evolved numerical simulation of a two-dimensional electron wave packet through a quantum double slit. J. Appl. Phys. {\bf 73}, 998 (1993).


\bibitem{cutler64} Cutler, P. H. \& Davis, J. C. Reflection and transmission of electrons through surface potential barriers. Surf. Sci. {\bf 1} 194-212 (1964).

\bibitem{lukes69} Lukes, T. \& Somaratna, T. S. The exact propagator for an electron in a uniform electric field and its application to Stark effect calculations. J. Phys. C (Solid St. Phys.) {\bf 2}, 586 (1969).

\bibitem{islam09} Islam, M. F. {\it et al.}, Solid State Comm. {\bf 149}, 1257 (2009).

\bibitem{fall99}Fall, C. Ab Initio Study of the Work Functions of Elemental Metal Crystal. Ph.D. thesis, \'{E}cole Polytechnique F\'{e}d\'{e}rale de Lausanne, (1999).


\end{thebibliography}
\end{document}